\documentclass[10pt,letterpaper]{IEEEtran}

\usepackage{graphicx}
\usepackage{amssymb}
\usepackage{amsmath}
\usepackage{amsfonts}
\usepackage{cite}
\usepackage{url}
\usepackage{enumerate}
\usepackage{epstopdf}
\usepackage{dsfont}
\usepackage{suffix}
\usepackage{steinmetz}
\usepackage{amsthm}
\usepackage{algpseudocode}
\usepackage{algorithm}

\makeatletter
\def\footnoterule{\relax%
  \kern-5pt
  \hbox to \columnwidth{\hfill\vrule width \columnwidth height 0.4pt\hfill}
  \kern4.6pt}
\makeatother

\begin{document}

\title{  \vspace{-0.1 in} Scatter Radio Receivers for Extended Range
Environmental Sensing WSNs} 
\vspace{-0.06 in}

\author{ \vspace{-0.08 in}
 Panos~N.~Alevizos and 
         Aggelos~Bletsas  
\\ 
\IEEEauthorblockA{School of Electronic and Computer Engineering, 
Technical University of Crete, Chania, Greece 73100\\
Email: palevizos$@$isc.tuc.gr, aggelos$@$telecom.tuc.gr }}

\maketitle
\vspace{-0.08 in}
\begin{abstract} 

Backscatter communication, relying on the reflection principle, 
constitutes a promising-enabling technology for
  low-cost, large-scale,  ubiquitous sensor networking. 
This work makes an overview of the state-of-the-art coherent and noncoherent
scatter radio receivers   that
 account for the peculiar  signal model
consisting of several microwave and communication parameters.

\end{abstract}

\vspace{-0.06 in}

\vspace{-0.06 in}
\section{Introduction}

The need of ubiquitous environmental sensing has lead to 
the adoption of cost-effective, large-scale
wireless sensor networks (WSNs). 
Such networks consists of several low-power and low-cost 
devices that sense 
 environmental variables
(such as  soil humidity, soil moisture, temperature) and gathering the sensed information at a central unit.
Existing commercial WSN equipment  incorporates devices consisting of
complex active {radio frequency} (RF) components that 
 increase significantly  the total  monetary cost, as well
as the overall energy consumption per sensor node.

Scatter radio adopts the  reflection principle \cite{Stockman:48}, 
which is achieved by generating a carrier wave that illuminates a set of  RF tag/sensors. A RF tag  terminates
its antenna load according to the data to be transmitted. 
The  incident signal is modulated and scattered back
towards a software-defined radio (SDR) reader for processing. 
The above idea is depicted in Fig.~\ref{fig:sys_mod}.

This work proposes bistatic scatter radio technology with semi-passive 
tags and frequency shift-keying (FSK) modulation, ideal for the power limited regime \cite{KiBlSa:14}.
 The specific three-fold design mixture, not only reduces  the energy cost per sensor tag,
 but also does not sacrifice
in total communication range, achieving ranges on the order of a hundred of meters, as opposed  to the conventional
passive radio frequency identification (RFID) Gen 2 architecture that offers a limited range on the order of very few meters.
A thorough overview of the state-of-the-art coherent and noncoherent
scatter radio FSK receivers is subsequently conducted. The receivers are capable to 
achieve one hundred and fifty meters communication range.

\vspace{-0.06 in}

\section{Bistatic Scatter Radio Signal Model}
\label{sec:Signal Model}

\begin{figure}[!t]
\centering
        \includegraphics[width=0.8\columnwidth]{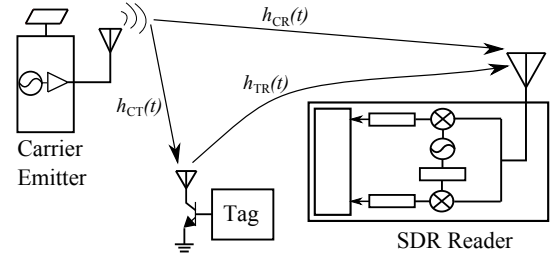}
\caption{Bistatic architecture system model: carrier emitter is displaced from
SDR reader and RF tag modulates the incident RF signal from carrier emitter.}
\label{fig:sys_mod}
\end{figure}

The bistatic scatter radio architecture is employed, consisting of
 a carrier emitter, a RF
tag, and a software-defined radio (SDR) reader
 (Fig.~\ref{fig:sys_mod}).
Due to the relatively small bit-rate (on the order of few kilobits per second), 
along with the fact that the carrier emitter-to-SDR reader  and  tag-to-SDR reader links  are on the order of
a kilometer, i.e., small delay spread,
 frequency  non-selective (flat) fading channel  is assumed, where the {baseband} complex channel  response
for a duration of channel coherence time, $T_{\rm coh}$, is given by
$
h_{\rm k}(t) = h_{\rm k} = a_{\rm k}  \mathsf{e}^{-\mathsf{j} \phi_{\rm k}} , ~ {\rm k}\in \{\rm CR,CT,TR\},
$
where $a_{\rm k},   \in \mathds{R}_+, ~{\rm k}\in \{\rm CR,CT,TR\},$  denote
 the channel attenuation parameters of the corresponding links and 
$\phi_{\rm k }  \in [0,2\pi),~ {\rm k}\in \{\rm CR,CT,TR\},$ stand for
 the respective phases due to signal propagation delay, all independent of each other.
Parameters $a_{\rm CR}, a_{ \rm CT}, a_{ \rm TR}$ are assumed
Rician distributed  due to strong line-of-site   signals, 
commonly encountered in outdoors WSNs.

Carrier emitter transmits a continuous sinusoid wave at carrier frequency $F_{\rm car}$ 
that illuminates the tag. Tag modulates the received signal at passband 
using  $50\%$ duty cycle square waveform pulse
of frequency $F_i$ and random initial phase $ \Phi_i \sim \mathcal{U}[0,2 \pi],~ i \in \mathds{B} \triangleq
\{0,1\}$ and   backscatters it towards SDR reader.
The received  complex baseband signal at the SDR reader is given
 by the superposition of the carrier emitter sinusoid and the backscattered
tag signal through channels $h_{\rm CR}$ and $h_{\rm TR}$, respectively.
The received signal also suffers from band-limited noise
with power spectral density $N_0/2$ over SDR receiver bandwidth $ W_{\rm SDR}$.
SDR reader applies  carrier frequency offset (CFO) estimation and compensation
using periodogram-based techniques, DC blocking, and  synchronization.

For $|F_0-F_1|\gg \frac{1}{T}$ and  $F_i \gg \frac{1}{T},$ the noise-free, CFO-free, DC-blocked, and perfect
synchronized baseband signal belongs to a four dimensional,   time limited  in $[0, T)$
signal space, whose orthonormal basis is denoted as
  $\mathcal{B}  \triangleq \left\{\frac{1}{\sqrt{T}}  \mathsf{e}^{\pm \mathsf{j} 2\pi F_i t}\Pi_T(t) 
 \right\}_{i \in \mathds{B}}$, where 
$\Pi_T(t)$ is the   square waveform of duration $T$.
The optimal demodulator projects
  received signal  on  basis $\mathcal{B}$ through correlators  and the
discrete baseband signal over bit duration $T$ can be written as \cite{FaAlBl:15} 
\begin{equation}
\vspace{-0.06 in}
\mathbf{r}
=
 \left[ \begin{array}{ccc} r_0^+ \\ r_0^- \\
r_1^+ \\ r_1^- \end{array} \right]  =
h \sqrt{\frac{\mathtt{E}}{2}}  \left[ \begin{array}{ccc}\mathsf{e}^{+ \mathsf{j} \Phi_0 } \\ \mathsf{e}^{- \mathsf{j} \Phi_0 }
\\ \mathsf{e}^{+ \mathsf{j} \Phi_1 } \\\mathsf{e}^{- \mathsf{j} \Phi_1 } \end{array} \right]
\odot
\mathbf{s}_i
+
\left[ \begin{array}{ccc} n_0^+ \\ n_0^- \\
n_1^+ \\ n_1^- \end{array} \right],
\label{eq:discrete_equivalent}
\end{equation}
where $h = \alpha_{\rm CT}\alpha_{\rm TR}\mathsf{e}^{-\mathsf{j} \phi}$ incorporates the fading  
coefficients $h_{\rm CT}h_{\rm TR} $ as well as 
the phase difference of compound link
carrier emitter-to-tag-to-SDR reader.
 $\mathtt{E}$ is the average energy per bit in baseband,
 $\mathbf{s}_i = [(1-i)~ (1-i)~ i~ i]^{\top}$ is the four-dimensional
symbol corresponding to bit $i \in \mathds{B}$,
and 
$
 \mathbf{n} = [n_0^+ ~n_0^- ~n_1^+ ~n_1^-]^{\top} \sim \mathcal{CN}\left(\mathbf{0}_4,
 \frac{N_0}{2} \mathbf{I}_4\right).
$

\vspace{-0.04 in}

\section{Bistatic Scatter Radio FSK Receivers}
\label{sec:Signal Model}

\subsection{Uncoded Reception}
\label{section:uncoded}
\subsubsection{Coherent Detector \cite{FaAlBl:15}}
The coherent receiver estimates the compound channel
  $\mathbf{h}_{c} \triangleq h \sqrt{\frac{\mathtt{E}}{2}}  \left[  \mathsf{e}^{+ \mathsf{j} \Phi_0 } ~
 \mathsf{e}^{- \mathsf{j} \Phi_0 }~ \mathsf{e}^{+ \mathsf{j} \Phi_1 } ~ \mathsf{e}^{- \mathsf{j} \Phi_1 }  \right]$
through the use of training symbols.
After obtaining the least-squares estimate of compound channel $\mathbf{h}_{c}$, 
$\widehat{\mathbf{h}}_{c}$, the  maximum-likelihood (ML)
detector  is applied through the rule \cite{FaAlBl:15}
\begin{equation}
\widehat{i}_{\rm ML} = \arg \max_{i \in \mathds{B}} \Re \!\left\{ 
\left(\widehat{\mathbf{h}}_{c} \odot \mathbf{s}_i \right)^{H} \mathbf{r}
\right\}.
\end{equation}

\subsubsection{Noncoherent Detectors \cite{AlBlKar:17}}

The first symbol-by-symbol noncoherent detector 
treats the parameter $h$ as deterministic and 
parameters $\{\Phi_i\}_{i \in \mathds{B} }$  as random; it is called hybrid composite
hypothesis testing (HCHT) detector and is given by \cite{AlBl:15}
\begin{align}
\vspace{-0.06 in}
&~\arg \max \limits_{i \in \mathds{B}} \left\{  \underset{\Phi_i}{\mathbb{E}}\!\left[
 \max \limits_{h \in \mathds{C}}  \mathsf{ln}\! \left( \mathsf{f}(\mathbf{r} |  i, h, \Phi_i) \right) \right]   \right\} 
\label{eq:hybrid_non_coh} \\
\Longleftrightarrow & ~ |r_0^+|^2 + |r_0^-|^2 
\underset{i=1}{\overset{i=0}{\gtrless}}
 |r_1^+|^2 + |r_1^-|^2 .
 \label{eq:sqare_law_detector}
\vspace{-0.02 in}
\end{align}
The second symbol-by-symbol
detector is the generalized-likelihood ratio test (GLRT)  detector, that  treats all unknown  parameters  as deterministic and 
is expressed as \cite{AlBlKar:17}
\begin{align}
\vspace{-0.06 in}
&~ \arg \max \limits_{i \in \mathds{B}}  \left\{ \max\limits_{\Phi_i \in [0,2\pi]}
 \max\limits_{h \in \mathds{C}}  \mathsf{ln}\!  \left( \mathsf{f}(\mathbf{r} |  i, h, \Phi_i) \right)   \right\} 
\label{eq:GLRT_symbol_by_symbol} \\
\Longleftrightarrow & ~
  |r_0^+| + |r_0^-| 
\underset{i=1}{\overset{i=0}{\gtrless}}
 |r_1^+| + |r_1^-| .
 \label{eq:sum_abs_detector}
\vspace{-0.02 in}
\end{align}
For a bit sequence of $N$ bits satisfying  $T_{\rm coh} \geq NT$,   the received sequence  a can be written as
$
  \mathbf{r}_{1:N} = \left[ \mathbf{r}_1 ~ \mathbf{r}_2 ~ \ldots ~
\mathbf{r}_N  \right] =  h \left[   \mathbf{x}_{i_1}\! (\Phi_{i_1}) ~
  \mathbf{x}_{i_2}\!(\Phi_{i_2})~\ldots ~
 \mathbf{x}_{i_N}\!(\Phi_{i_N})  \right] + \left[ \mathbf{n}_1  ~\mathbf{n}_2 ~ \ldots ~
\mathbf{n}_N  \right],$ with 
$\mathbf{x}_{i_n} \! (\Phi_{i_n}) \triangleq  \sqrt{\frac{\mathtt{E}}{2}} \left[ \mathsf{e}^{\mathsf{j}\Phi_0}~\mathsf{e}^{-\mathsf{j} \Phi_0}~
\mathsf{e}^{\mathsf{j}\Phi_1}~ \mathsf{e}^{-\mathsf{j}\Phi_1} \right]^{\top} \odot  \mathbf{s}_{i_n},$
with $i_n \in \mathds{B}$,
$n=1,2,\ldots, N$.  To reduce the sequence error rate, a noncoherent sequence detector
may be applied. For the above signal model, the GLRT sequence detector is expressed as  
\begin{equation}
\arg \max\limits_{\mathbf{i} \in  \mathds{B}^N} \left\{ \max\limits_{({\Phi}_0, \Phi_1) \in [0,2\pi]^2}
 \max\limits_{h \in \mathds{C}}  \mathsf{ln}\! \left( \mathsf{f}\! \left (\mathbf{r}_{1:N} | \mathbf{i} , h, 
{\Phi}_0, \Phi_1 \right)  \right) \right\}.
\end{equation}
Work in  \cite{AlBlKar:17} partitions the space of
phases $({\Phi}_0, \Phi_1)$ in distinct $M^2$ points and solves the above problem 
with complexity $\mathcal{O}(N\mathsf{log}(N))$ using the
same procedure  with the work in  \cite{AlFouKarBl:16}.

\subsection{Coded Reception}
\label{section:coded}
When channel coding is utilized, the transmitter
encodes a sequence of $K$ bits to a sequence
of $N$ coded bits, $\mathbf{c}=[c_1~c_2~\ldots,c_N]$ belonging to a
code $\mathcal{C} \subset \mathds{B}^N$. 

\subsubsection{Coherent Decoder \cite{FaAlBl:15}}

After obtaining the least-squares estimate of compound channel $\mathbf{h}_{c}$, 
$\widehat{\mathbf{h}}_{c}$, the  ML decoder 
can be expressed as \cite{FaAlBl:15}
\begin{equation}
\vspace{-0.06 in}
\widehat{\mathbf{c}}_{\rm ML} = \arg \max_{\mathbf{c} \in \mathcal{C}} \sum_{n=1}^N \Re \!\left\{ 
\left(\widehat{\mathbf{h}}_{c} \odot \mathbf{s}_{c_n} \right)^{H} \mathbf{r}_n
\right\}
\end{equation}
It can be shown that, if the above decoder is combined  with interleaving
technique,  it can achieve 
diversity order $d_{\rm min}$, where $d_{\rm min}$
is the minimum distance of code $\mathcal{C}$ \cite{FaAlBl:15}.

\vspace{-0.02 in}

\subsubsection{Nonoherent Decoder  \cite{AlBlKar:17}}

Suppose that we employ channel coding, noncoherent receiver, and the use of interleaving technique  of
 interleaving depth $D$ 
satisfying $D T \geq T_{\rm coh}$; accordingly with Eq.~\eqref{eq:hybrid_non_coh},   HCHT decoder 
can be derived as \cite{AlBl:15}
\begin{equation}
\vspace{-0.05 in}
  \widehat{\mathbf{c}} =   \arg \max_{\mathbf{c} \in \mathcal{C}} \sum_{n=1}^N 
\left(\left\|\mathbf{r}_n \odot \mathbf{s}_{1}\right\|^2-(\left\|\mathbf{r}_n \odot \mathbf{s}_{0}\right\|^2
\right)  c_n .
 \label{eq:comp_hyp_decoding rule}
\end{equation}
For the case of $D T< T_{\rm coh}$, the above decoder is suboptimal,
however, it is utilized for any value of $D$, due to its inherent simplicity.

\vspace{-0.07 in}
\section{Simulation Results}
\label{sec:simul_Results}

Simulations are conducted for bistatic scatter radio system model
assuming backscatter  FSK transmissions with bit rate $T=1$msec, over quasi-static
Rician flat fading channel with coherence time $T_{\rm coh}=100$msec.
In all simulation results we have assumed  perfect  
synchronization, perfect CFO estimation/compensation,
and  fixed energy budget per transmitted packet for all schemes.


\begin{figure}[!t]
\centering
\vspace{-0.1 in}
        \includegraphics[width=0.76\columnwidth]{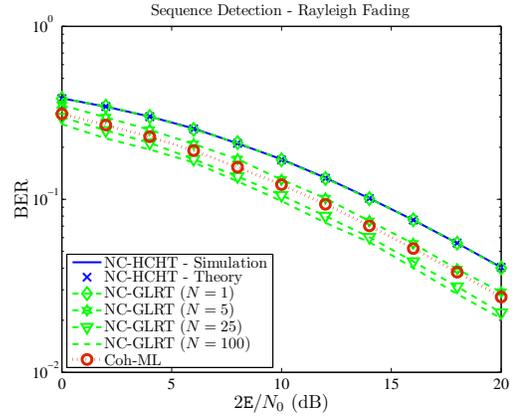}
\caption{BER   vs average received SNR  for  uncoded   schemes. Due to fixed energy budget per transmitted packet and the fact that
coherent receiver employees extra $30$ bits for channel estimation (i.e., it 
has less energy per bit), noncoherent GLRT sequence
 detector outperforms coherent one.}
\label{fig:NonCohSequence_vs_Coh_ImpacSequenceLength}
\end{figure}

%

Fig.~\ref{fig:NonCohSequence_vs_Coh_ImpacSequenceLength} 
illustrates the bit error rate (BER)   performance of all uncoded schemes studied in Section~\ref{section:uncoded}
as function of average received signal-to-noise ratio, $\frac{2\mathtt{E}}{N_0}$.
 We observe that GLRT sequence
detector  outperforms all  schemes. It is  also noted that coherent ML
detector offers 3dB better BER performance compared to noncoherent symbol-by-symbol schemes. 

\vspace{-0.05 in}
\bibliographystyle{IEEEtran}
\bibliography{IEEEabrv,BistaticBib_v2}

\end{document}